\documentclass{article}
\usepackage{icmc2026_paper_template}
\usepackage{times}
\usepackage{ifpdf}
\usepackage{soul}
\usepackage{amsmath}
\usepackage{amssymb}
\usepackage[english]{babel}

\def\papertitle{Vibrato Matching for Modulation Control and Blending in Sound Mixtures}
\def\firstauthor{Jeremy Hyrkas}

\newif\ifpdf
\ifx\pdfoutput\relax
\else
   \ifcase\pdfoutput
      \pdffalse
   \else
      \pdftrue
  \fi
\fi

\ifpdf 
  \usepackage[pdftex,
    pdftitle={\papertitle},
    pdfauthor={\firstauthor},
    bookmarksnumbered, 
    pdfstartview=XYZ 
   ]{hyperref}

  \usepackage[pdftex]{graphicx}
  \graphicspath{{./figures/}}
  \DeclareGraphicsExtensions{.pdf,.jpeg,.png}

  \usepackage[figure,table]{hypcap}

\else 
  \usepackage[dvips,
    bookmarksnumbered, 
    pdfstartview=XYZ 
  ]{hyperref}  

  \usepackage[dvips]{epsfig,graphicx}
  \graphicspath{{./figures/}}
  \DeclareGraphicsExtensions{.eps}

  \usepackage[figure,table]{hypcap}
\fi

\hypersetup{
    colorlinks,%
    citecolor=black,%
    filecolor=black,%
    linkcolor=black,%
    urlcolor=black
}

\title{\papertitle}

\oneauthor
   {\firstauthor} {University of California San Diego \\ %
     {\tt \href{mailto:jhyrkas@ucsd.edu}{jhyrkas@ucsd.edu}}}

\begin{document}
\capstartfalse
\maketitle
\capstarttrue
\begin{abstract}
In sound mixtures of more than one musical source, different vibrato patterns act as a cue that multiple sources are present for both human listeners and source separation algorithms.
Matching the vibrato patterns of the signals in the mixture reduces the perception of multiple sources, particularly when the sources play in unison.
This work introduces the vibrato matching algorithm, which first suppresses vibrato in a target signal and then transfers vibrato from a source signal to the target.
An existing vibrato suppression algorithm is combined with a new algorithm for vibrato transfer, which imparts frequency modulation and amplitude modulation to the harmonics of the target signal, and amplitude modulation onto the spectral envelope of the non-harmonic residual component.
Examples demonstrate the algorithm's utility as a vibrato control mechanism and as a tool for blending sound sources.
Matching vibrato degrades the performance of source separation algorithms, suggesting a similar degradation in listeners ability to detect the presence of multiple sources.
\end{abstract}

\section{Introduction}
\label{sec:introduction}

Vibrato is a musical technique in which the pitch of a sustained note is modulated, and is often accompanied by a modulation in sound level.
Signals with natural vibrato therefore feature both frequency and amplitude modulation (FM/AM). 
The presence of different vibrato patterns improves the ability of listeners to count the number of instruments in a mix, even when the instruments play in unison~\cite{Stoter2013,Schoeffler2013}.
This finding motivated the development of unison source separation algorithms that use modulation patterns to isolate sources~\cite{Stoter2014,Stoter2016}.

Given its perceptual salience, precise control of a signal's vibrato can be useful.
Vibrato can be manipulated using sinusoidal modeling~\cite{Roebel2011}, where the AM and FM of harmonic partials and AM in the spectral envelope are analyzed and altered prior to resynthesis.
Recent efforts focus on manipulating signals directly, either to remove existing vibrato~\cite{Hyrkas2025ISMRA} (\emph{vibrato suppression}) or to apply vibrato patterns from one signal to another~\cite{Hyrkas2024DAFx,Hyrkas2025ICMC} (\emph{vibrato transfer}).

This study combines and expands these approaches to introduce the \emph{vibrato matching} algorithm.
Vibrato suppression~\cite{Hyrkas2025ISMRA} is reviewed in Section~\ref{sec:suppression}.
Section~\ref{sec:transfer} proposes an update to the vibrato transfer algorithm~\cite{Hyrkas2025ICMC} that expands its AM transfer.
Vibrato matching, the combination of vibrato suppression and transfer, is evaluated in Section~\ref{sec:matching} as both a means to change the vibrato pattern of a signal and as a tool for matching vibrato patterns in sound mixtures.
It is also shown that by vibrato matching reduces the quality of unison source separation.
This reduction suggests that matching vibrato patterns of unison signals can be used as a creative tool to obscure the presence of multiple instruments in a sound mixture.

\section{Vibrato Suppression}
\label{sec:suppression}
The goal of vibrato matching is to analyze a source signal with vibrato $s(n)$ and to impart its modulation patterns to a target signal $t(n)$.
Any existing vibrato in $t(n)$ should be suppressed before applying new vibrato to avoid interference of the modulation patterns.
Previously proposed vibrato suppression methods are briefly reviewed below and used as a pre-processing step to vibrato transfer.

\subsection{Vibrato Filter}
\label{sec:vib_filter}
Previous works on vibrato control use a \emph{vibrato filter}~\cite{Roebel2011,Hyrkas2025ISMRA} $v_f(n)$ in various stages of analysis and synthesis.
The vibrato filter is a high-order FIR lowpass filter with a cutoff frequency lower than typical vibrato rates (i.e., 2 Hz).
The filter is applied using forward-backward filtering to create a zero-phase filter.
This filter is used on frequency or amplitude envelopes to suppress variations caused by vibrato.

\subsection{Suppressing FM caused by vibrato}
\label{sec:suppress_fm}

\begin{figure*}
\centering
\includegraphics[width=\linewidth]{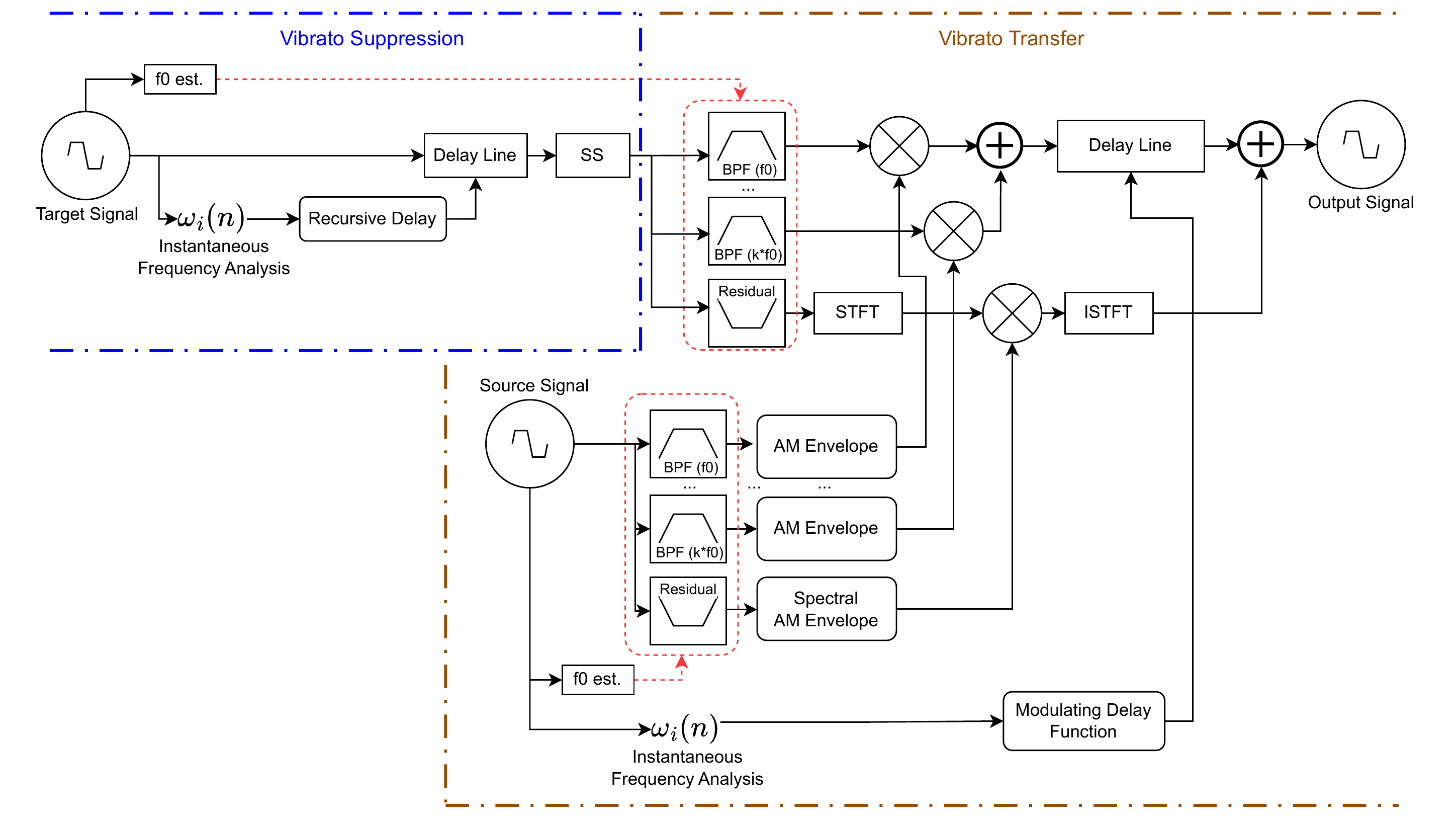}
\caption{Signal flow for the vibrato matching algorithm. 
A target signal $t(n)$ has its original vibrato suppressed and is then modulated using AM and FM patterns from a vibrato source signal $s(n)$.
}
\label{fig:matching_alg}    
\end{figure*}

The signal $\omega_i^t(n)$ contains the instantaneous fundamental frequency of $t(n)$.
$\omega_i^t(n)$ can be derived using peak-picking in the short-time Fourier transform (STFT) of $t(n)$ or from the analytic signal of the first harmonic isolated using a bandpass filter~\cite{Hyrkas2024DAFx}.
\begin{equation}
    \overline{\omega}_i^t(n) = \omega_i^t(n) \circledast v_f(n)
    \label{eq:wit_vf}
\end{equation}
is the instantaneous fundamental frequency of $t(n)$ after removing frequency deviations caused by vibrato~\cite{Hyrkas2025ISMRA}.

Using these signals, a time-varying demodulating delay
\begin{equation}
    D_d(n) = \sum_{l=0}^n \Big( \frac{\omega_i^t(l)}{\overline{\omega}_i^t(l)} - 1 \Big)
    \label{eq:Dd}
\end{equation}
is constructed. 
The signal $t\big(n-D_d(n)\big)$ contains very little pitch modulation when $t(n)$ contains light vibrato.
For heavier vibrato extent, the \emph{recursive delay} 
\begin{equation}
    D_{dd}(n) = D_d\big(n-D_d(n)\big)
    \label{eq:Ddd}
\end{equation}
is more effective.
The signal
\begin{equation}
    \widehat{t}(n) = t\big(n-D_{dd}(n)\big)
    \label{eq:t_hat}
\end{equation}
contains no audible FM, but may still contain AM~\cite{Hyrkas2024DAFx}.

\subsection{Suppressing AM caused by vibrato}
\label{sec:suppress_am}
To suppress AM in $\widehat{t}(n)$, \emph{spectrogram smoothing} (SS)~\cite{Hyrkas2025ISMRA} is performed.
\begin{equation}
    \widehat{T}(\omega,m) = \mathrm{STFT}\big(\widehat{t}(n)\big)
    \label{eq:T_hat}
\end{equation}
is used to derive
\begin{equation}
    E_T(\omega,m) = |\widehat{T}(\omega,m)| \; ,
    \label{eq:ET}
\end{equation}
the magnitude spectrogram of $\widehat{t}(n)$.
A vibrato filter is applied along the $m$ axis for each frequency $\omega$ to derive
\begin{equation}
    \overline{E}_T(\omega,m) = E_T(\omega,m) \circledast v_f(m) \; ,
    \label{eq:ET_vf}
\end{equation}
the magnitude spectrogram without vibrato-induced AM.
Finally, the ratio of these spectrograms is used to derive
\begin{equation}
    \overline{t}(n) = \mathrm{ISTFT}\Big((\widehat{T}(\omega,m) \cdot \frac{\overline{E}_T(\omega,m)}{E_T(\omega,m)}\Big) \; ,
    \label{eq:SS}
\end{equation}
a version of the target signal with no vibrato-induced FM or AM.
Vibrato suppression using the recursive delay and SS is depicted in the upper left portion of Figure~\ref{fig:matching_alg}.

\section{Vibrato Transfer}
\label{sec:transfer}

The FM patterns from a source signal with vibrato $s(n)$ can be transferred to the vibrato-suppressed target signal $\overline{t}(n)$ using a time-varying delay line.
Previous work performed this transfer in real-time with the addition of a simple AM envelope~\cite{Hyrkas2025ICMC}. 
In real acoustic vibrato, AM patterns are often different in each harmonic~\cite{Zhang2015} and across the spectral envelope of the non-harmonic \emph{residual} portion of the signal~\cite{Roebel2011}.
Realistic vibrato transfer therefore requires modulating the amplitudes and frequencies of the harmonics of $\overline{t}(n)$ and applying AM to the residual signal.

\subsection{Transferring AM to the harmonics}
\label{sec:am_harms}

Vibrato transfer begins by deriving AM envelopes from the harmonics of $s(n)$ and applying them to corresponding harmonics in $\overline{t}(n)$.
Using the fundamental frequency $f_0^s$ of $s(n)$\footnote{Known a priori or determined using an $f_0$ detection algorithm},
the $k$th harmonic of $s(n)$ is isolated using
\begin{equation}
    h_k^s(n) = s(n) \circledast \mathrm{BP}_k^s(n) \; ,
    \label{eq:s_bp}
\end{equation}
where $\mathrm{BP}_k^s(n)$ is a Butterworth bandpass filter centered at frequency $k \cdot f_0^s$.

\begin{figure}
\centering
\includegraphics[width=\linewidth]{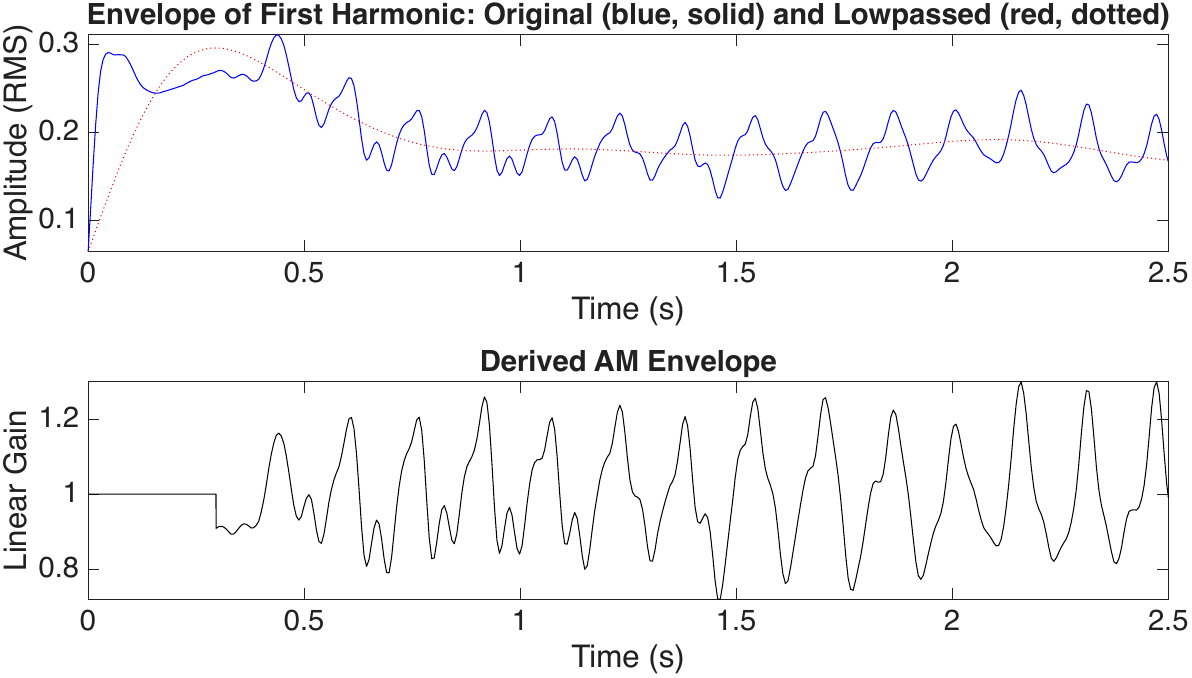}
\caption{Deriving an AM envelope from an RMS envelope.}
\label{fig:am_env}    
\end{figure}

The root-mean-square (RMS) envelope of each harmonic is computed as
\begin{equation}
    E_{k}^s(n) = \mathrm{RMS}\big(h_k^s(n)\big) 
    \label{eq:s_rms}
\end{equation}
over sliding windows and interpolated to signal length.
\begin{equation}
    \overline{E}_{k}^s(n) = E_{k}^s(n) \circledast v_f(n)
    \label{eq:s_rms_vf}
\end{equation}
is the envelope of $h_k^s(n)$ without vibrato-induced AM.
\begin{equation}
    \mathrm{AM}_k (n) = \frac{E_{k}^s(n)}{\overline{E}_{k}^s(n)}
    \label{eq:amk}
\end{equation}
yields the AM envelope of the $k$th harmonic 
(i.e., the time-varying amplitude such that $E_{k}^s(n) = \mathrm{AM}_k(n) \cdot \overline{E}_{k}^s(n)$).
It is useful to set $\mathrm{AM}_k(n)=1$ until the first peak in $\overline{E}_{k}^s(n)$ so that the signal's attack is preserved.
Figure~\ref{fig:am_env} shows an AM envelope derived from a corresponding RMS envelope.

With AM envelopes derived for each harmonic in $s(n)$, the harmonics of $\overline{t}(n)$ are now isolated as
\begin{equation}
    h_k^t(n) = \overline{t}(n) \circledast \mathrm{BP}_k^t(n) \; ,
    \label{eq:t_bp}
\end{equation}
where $\mathrm{BP}_k^t(n)$ is a Butterworth bandpass filter centered at frequency $k \cdot f_0^t$.
Each harmonic is amplitude modulated
\begin{equation}
    \widetilde{h_k^t}(n) = h_k^t(n) \cdot \mathrm{AM}_k(n) \; ,
    \label{eq:t_bp_star}
\end{equation}
after which the amplitude modulated harmonics are summed to create
\begin{equation}
    \widetilde{h_t}(n) = \sum^k \widetilde{h_k^t}(n) \; .
    \label{t_bp_star_sum}
\end{equation}

\subsection{Transferring FM to the harmonics}
\label{sec:fm_harms}
As in Section~\ref{sec:suppress_fm}, $\omega_i^s(n)$ contains the instantaneous fundamental frequency of $s(n)$ and
\begin{equation}
    \overline{\omega}_i^s(n) = \omega_i^s(n) \circledast v_f(n)
    \label{eq:wis_vf}
\end{equation}
is the instantaneous fundamental frequency without vibrato.

\begin{equation}
    D_m(n) = \sum_{l=0}^n \Big( 1 - \frac{\omega_i^s(l)}{\overline{\omega}_i^s(l)} \Big)
    \label{eq:Dm}
\end{equation}
is a time-varying modulating delay that causes the same FM pattern present in $s(n)$~\cite{Hyrkas2025ISMRA}.
Using this delay function on $\widetilde{h_t}$ yields
\begin{equation}
    h_t^*(n) = \widetilde{h_t}\big(n - D_m(n) \big) \; ,
    \label{eq:h_star}
\end{equation}
a signal comprised of the harmonics of $t(n)$ with identical FM and AM pattern as the harmonics of $s(n)$.

\subsection{Transferring AM to the residual signal}
\label{sec:am_noise}

AM in the non-harmonic residual in signals with vibrato can differ across the frequency spectrum~\cite{Roebel2011}, so it is best imparted in a frequency-dependent manner.

The residual signal of $s(n)$ is captured as
\begin{equation}
    r_s(n) = s(n) \circledast \mathrm{BS}_1^s(n) \circledast \mathrm{BS}_2^s(n) \circledast \ldots \circledast \mathrm{BS}_k^s(n)\; ,
    \label{eq:r_s}
\end{equation}
where each $\mathrm{BS}_k^s(n)$ is a Butterworth bandstop filter centered at frequency $k \cdot f_0^s$.
$\mathrm{BS}_k^s(n)$ rejects the $k$th harmonic in $s(n)$ and is complementary to $\mathrm{BP}_k^s(n)$ in Section~\ref{sec:am_harms}.

The STFT of the residual signal
\begin{equation}
    R_s(\omega,m) = \mathrm{STFT\big(r_s(n)\big)}
    \label{eq:R_s}
\end{equation}
is divided into $d$ equally spaced and non-overlapping frequency regions across $\omega$.
For each frame $m$, the frequency with maximum magnitude in each region
\begin{equation}
    \omega_d^*(m) = \underset{\omega*\in[\omega_d,\omega_{d+1})}{\arg\max} |R_s(\omega^*,m)|
    \label{eq:omega_dstar}
\end{equation}
is used to create the spectral envelope
\begin{equation}
    \mathrm{SE}_{\mathrm{dB}}(d,m) =  20 \log_{10}\Big( |R_s\big(\omega_d^*(m),m\big)| \Big)
    \label{eq:se_dB}
\end{equation}
in decibels. A vibrato filter is applied across the $m$ dimension for each region to create 
\begin{equation}
    \overline{\mathrm{SE}}_{\mathrm{dB}}(d,m) = \mathrm{SE}_{\mathrm{dB}} \circledast v_f(m) \; ,
    \label{eq:se_dB_vf}
\end{equation}
the spectral envelope without vibrato-induced AM.

The spectral envelopes are converted to linear amplitude
\begin{equation}
    \mathrm{SE}_s(d,m) = 10^{\mathrm{SE}_{\mathrm{dB}}(d,m) \cdot \frac{1}{20}} 
    \label{eq:se_l}
\end{equation}
\begin{equation}
    \overline{\mathrm{SE}}_s(d,m) = 10^{\overline{\mathrm{SE}}_{\mathrm{dB}}(d,m) \cdot \frac{1}{20}} \; ,
    \label{eq:se_vf_l}
\end{equation}
and a spectral AM envelope is constructed as
\begin{equation}
    \mathrm{AM}_s(d,m) = \frac{\mathrm{SE}_s(d,m)}{\overline{\mathrm{SE}}_s(d,m)}.
\end{equation}
Finally, $\mathrm{AM}_s(d,m)$ is linearly interpolated across $d$ back to the original cardinality of $R_s(\omega,m)$ to create $\mathrm{AM}_s(\omega,m)$. 

The residual of $\overline{t}(n)$ is isolated using
\begin{equation}
    r_t(n) = \overline{t}(n) \circledast \mathrm{BS}_1^t(n) \circledast \mathrm{BS}_2^t(n) \circledast \ldots \circledast \mathrm{BS}_k^st(n)\; ,
    \label{eq:r_t}
\end{equation}
where each $\mathrm{BS}_k^t(n)$ is a Butterworth bandstop filter centered at frequency $k \cdot f_0^t$.
The spectral envelope of $r_t(n)$ is modulated using the spectral AM envelope as
\begin{equation}
    r_t^* = \mathrm{ISTFT}\Big( \mathrm{STFT}\big( r_t(n) \big) \cdot \mathrm{AM}_s(\omega,m) \Big) \; .
    \label{eq:r_t_star}
\end{equation}
Finally, the modulated residual is added with the modulated harmonics to create
\begin{equation}
    t^*(n) = h_t^*(n) + r_t^* \; ,
    \label{eq:t_star}
\end{equation}
the target $t(n)$ with its vibrato matched to the source $s(n)$.
The signal flow of the vibrato transfer portion of the vibrato matching algorithm is depicted on the bottom and right portions of Figure~\ref{fig:matching_alg}.

\section{Applications of Vibrato Matching}
\label{sec:matching}

Potential applications of vibrato matching are showcased here, both as a tool for modifying the vibrato of a target signal and as a method of controlling modulation in mixtures of signals.
Accompanying audio, as well as supplemental additional figures and sound examples from each section, are available on
\href{https://jeremyhyrkas.com/ICMC2026}{the supplemental website}.

\subsection{Modifying Vibrato in a Target Signal}
\label{sec:modifying}
Vibrato matching allows a target signal's vibrato to be precisely matched to a source.
This modification may be useful when the target's vibrato is unsatisfactory, which could be caused by a poor recorded performance 
or the use of a synthesis technique that lacks realistic vibrato.

\begin{figure*}
\centering
\includegraphics[width=\linewidth]{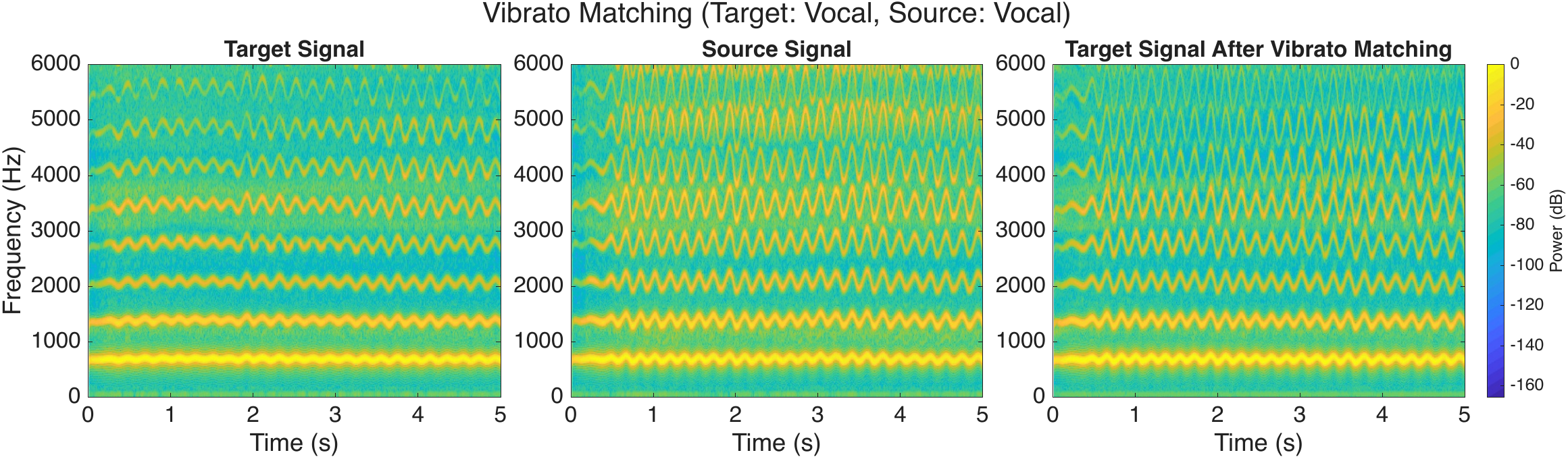}
\caption{Spectrograms of a vocal with vibrato (left), a vocal with different vibrato (middle), and the left signal after its vibrato was suppressed and vibrato from the middle signal was transferred onto it (right).}
\label{fig:before_and_after1}    
\end{figure*}

\begin{figure*}
\centering
\includegraphics[width=\linewidth]{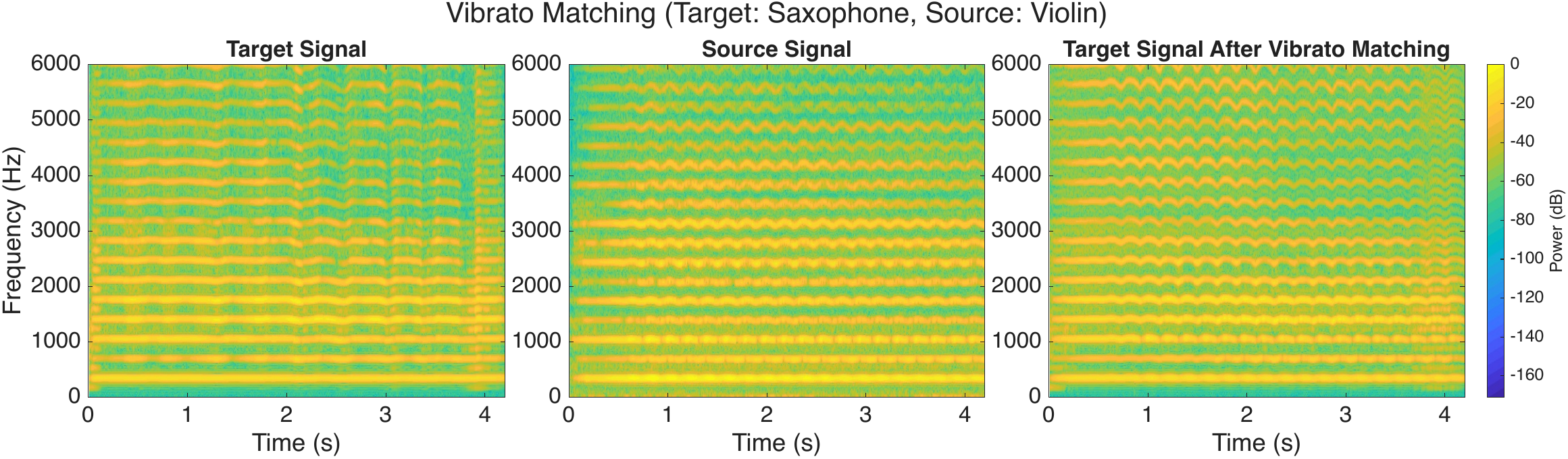}
\caption{Spectrograms of a saxophone with vibrato (left), a violin with vibrato (middle), and the saxophone after its vibrato was suppressed and vibrato from the violin was transferred onto it (right).}
\label{fig:before_and_after2}    
\end{figure*}

Figure~\ref{fig:before_and_after1} shows vibrato matching applied to vocal signals, in which the source signal has a higher vibrato extent than the target.
The rate and depth of vibrato in the target are matched to those in the source.
Figure~\ref{fig:before_and_after2} depicts vibrato matching between instrument types.
A saxophone signal has its vibrato matched to that of a violin, creating the sound of a woodwind instrument with string-like vibrato.

\subsection{Modulation Control in Sound Mixtures}
\label{sec:mixtures}

\begin{figure}
\centering
\includegraphics[width=\linewidth]{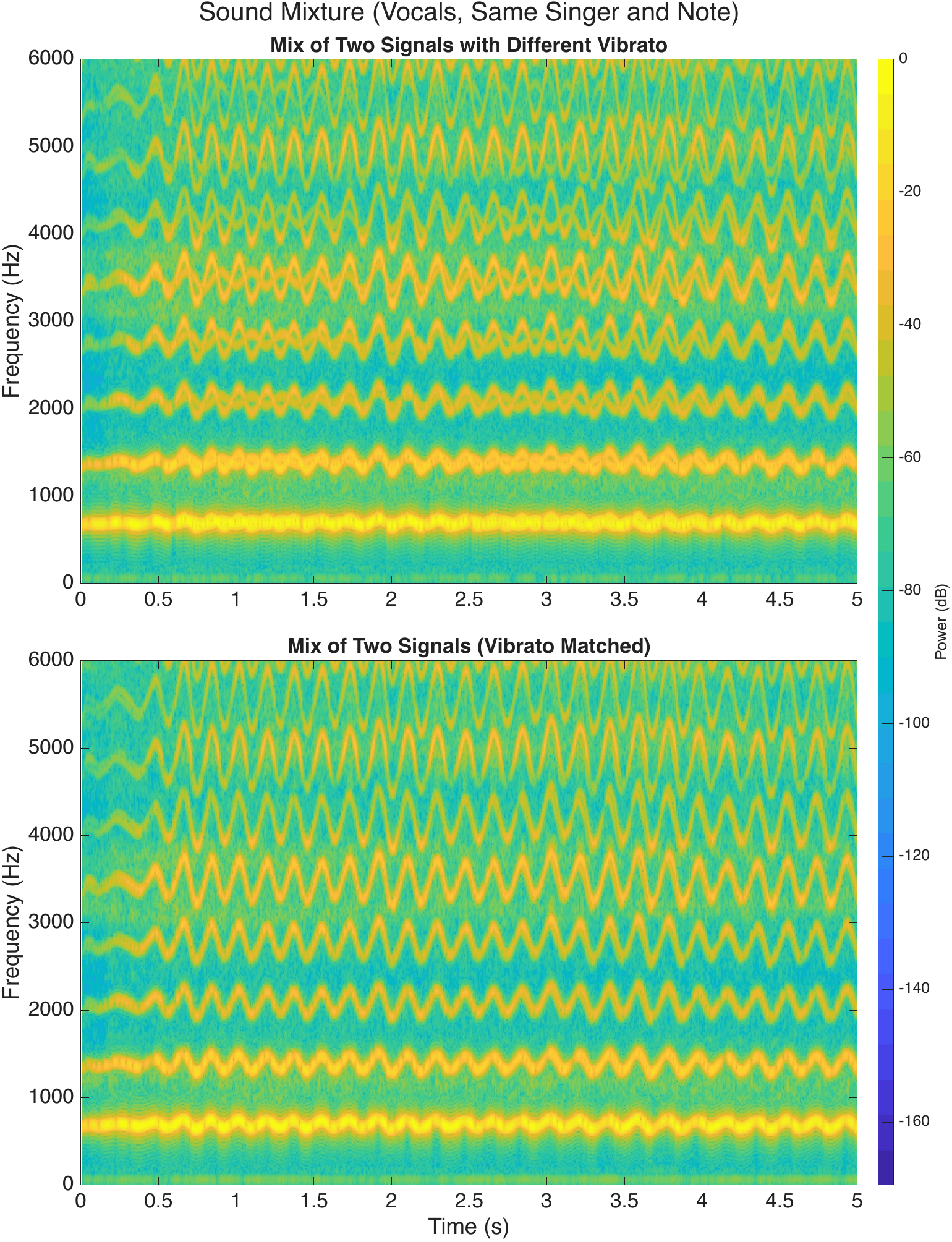}
\caption{Spectrograms of two vocal signals in unison, before and after matching vibrato patterns.}
\label{fig:matching_natnat}    
\end{figure}

Vibrato matching can be used when mixing signals with vibrato as both an engineering tool and a creative audio effect.
Figure~\ref{fig:matching_natnat} demonstrates a mix of the source and target vocal signals from Figure~\ref{fig:before_and_after1}.
In the top spectrogram, partials from each source are visible due to their different modulation patterns.
Sonically, the mix contains rough beating as the signals fall in and out of tune with each other.

After vibrato matching, the spectrogram appears to contain only one signal even though two are present.
The vibrato matched mix sounds similar to a double tracked vocal performance with no inconsistencies in vibrato pattern between takes.
This example demonstrates vibrato matching's usefulness as an audio engineering tool when double tracking performances with vibrato.

\begin{figure}
\centering
\includegraphics[width=\linewidth]{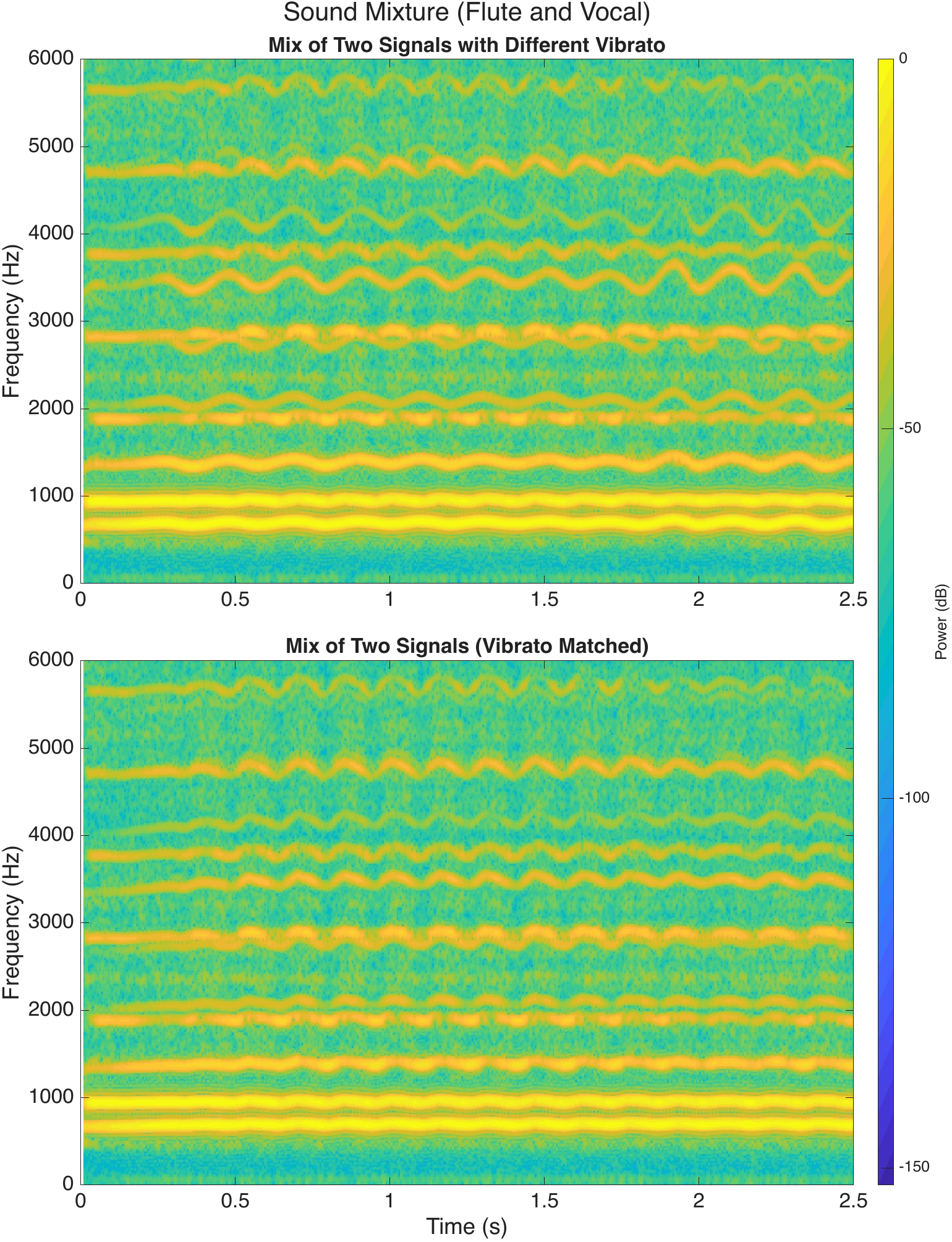}
\caption{Spectrograms of a vocal signal and a flute signal separated by a perfect fifth, before and after matching vibrato patterns.}
\label{fig:matching_flutenat}    
\end{figure}

Figure~\ref{fig:matching_flutenat} depicts a more creative approach.
The vibrato pattern from a flute is transferred to a vocal signal, imparting AM in the residual noise that is uncommon in vocal vibrato.
In this example, vibrato matching is used as a sound design tool to influence a listener's perception of sonic grouping.
When the vibrato matched signals are mixed, the signals modulate parallel, creating a sense of shared movement\footnote{
Grouping elements based on the perception of shared motion is described by the Gestalt principle of Common Fate. The Common Fate Transform used in unison source separation is named for this principle.} 
even though two sources are perceived due to their different pitches and timbre.

\subsection{Sound Blending and Source Separation}
\label{sec:blending}

Different vibrato patterns aid in detecting the presence of multiple sources in a mix for both listeners and unison source separation algorithms.
If matching the vibrato patterns of unison signals degrades the quality of source separation, it stands to reason that it may also degrade a listener's ability to perceive multiple instruments.
Vibrato matching could therefore be used as a sound blending tool in which multiple instrument sounds are mixed and perceived as one sound source.

To test the effect of vibrato matching on source detection, the source separation algorithm based on the Common Fate Transform (CFT)~\cite{Stoter2016} was used on several examples of two unison signals before and after vibrato matching.
This algorithm was previously shown to be highly effective at separating unison signals based on their FM and AM patterns.
Each signal analyzed by the algorithm contains three events: one source in isolation, then the second source in isolation, and finally both sources simultaneously.
This sequencing follows examples from the CFT's original evaluation in an attempt to maximize the separation quality.

\begin{figure}
\centering
\includegraphics[width=\linewidth]{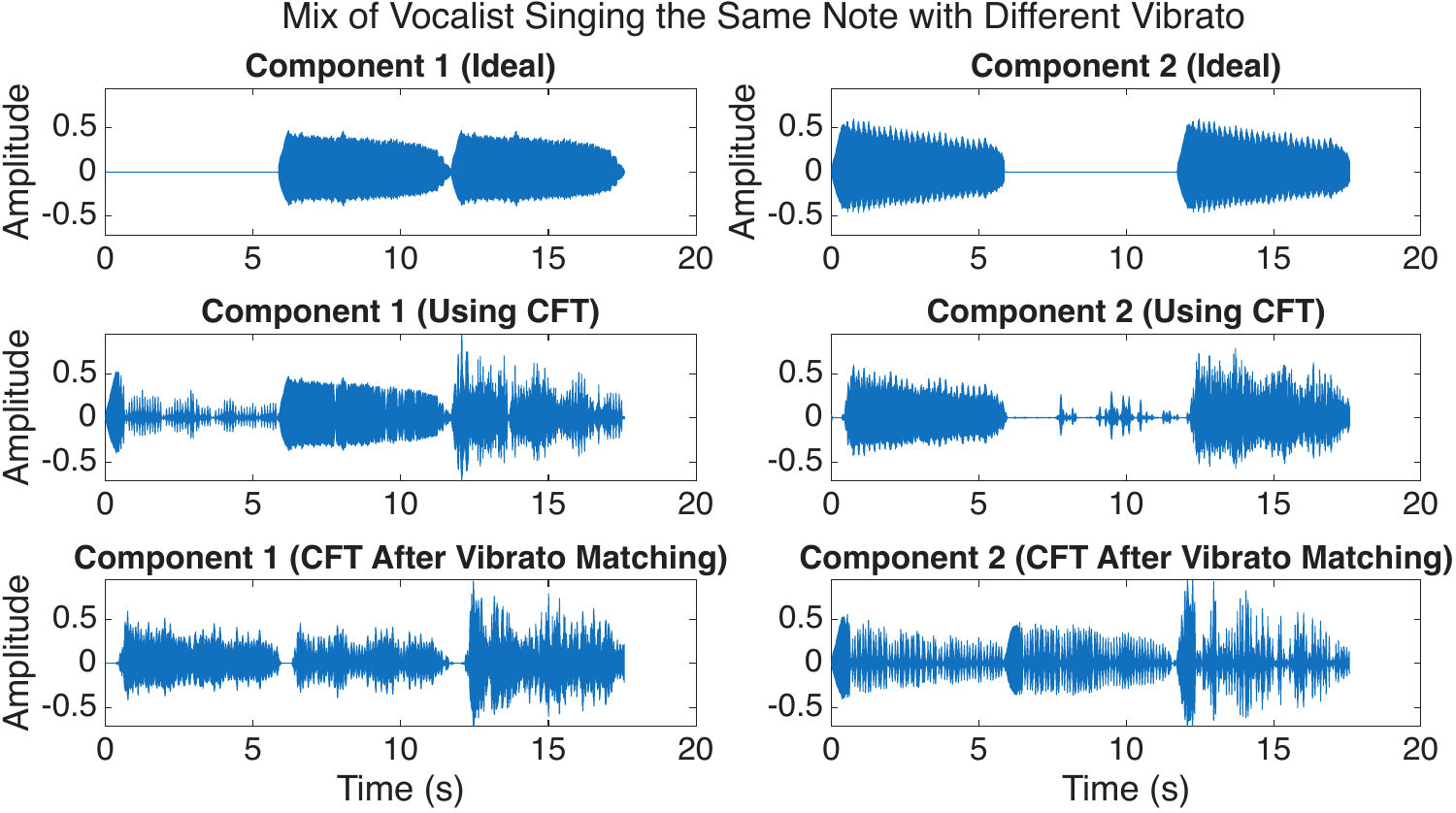}
\caption{Unison source separation on a mixture of two vocal signals, before and after matching vibrato patterns.}
\label{fig:separation_voxvox}    
\end{figure}

Figure~\ref{fig:separation_voxvox} shows source separation of the vocal mixes from Figure~\ref{fig:matching_natnat}.
The mixes feature a vocalist singing the same note with different vibrato rates.
The ideal outputs of the source separation algorithm are shown in the top row.
The middle row demonstrates the effectiveness of source source separation using the CFT. 
The first source has noticeable leakage, while the second source is well isolated.
The bottom row shows that vibrato matching completely degrades the CFT's ability to separate these sources.
This result is congruent with the bottom spectrogram in Figure~\ref{fig:separation_voxvox}, in which the two sources visually appear as one.

\begin{figure}
\centering
\includegraphics[width=\linewidth]{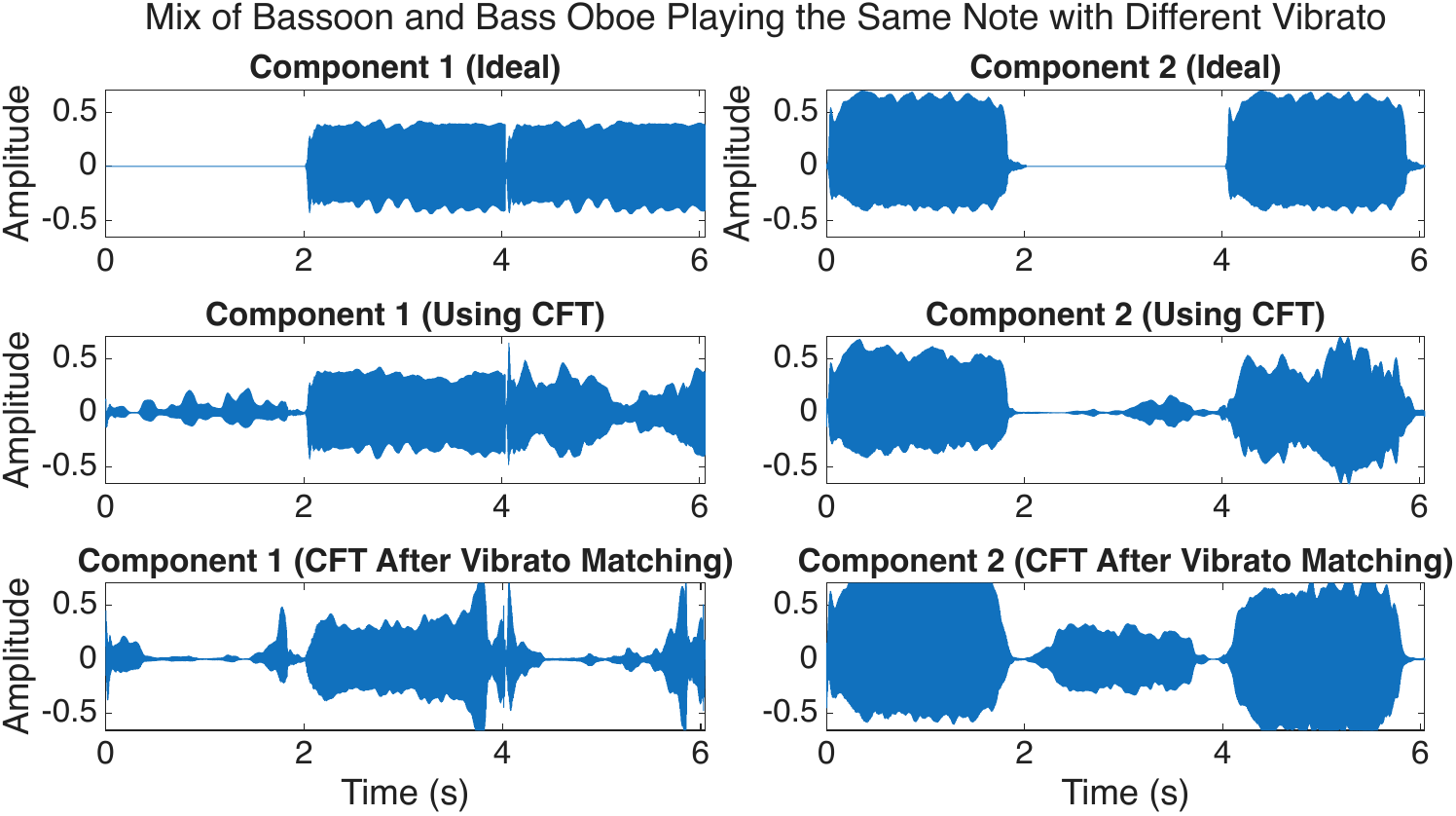}
\caption{Unison source separation on a mixture of bass oboe and bassoon, before and after matching vibrato patterns.}
\label{fig:separation_oboebassoon}    
\end{figure}

A mix of bassoon and bass oboe playing in unison is shown in Figure~\ref{fig:separation_oboebassoon}.
The bottom row again shows that vibrato matching reduces the effectiveness of source separation when compared to the original mix in the middle row.
The second source is split between the two separated components (see the middle third of each component). 
It is clear from the final third of each component that the second component extracted from the vibrato matched mix contains the majority of each source's signal instead of only one source.
This indicates that the separation algorithm cannot unmix the sources even when primed with each source in isolation.

Spectral features other than modulation play a role in unison source separation.
Nevertheless, Figures~\ref{fig:separation_voxvox} and~\ref{fig:separation_oboebassoon} indicate a reduced ability for a machine (and perhaps a human) listener to detect the presence of multiple sources in a unison mix after vibrato patterns in each source are matched.

\section{Conclusion}
\label{sec:conclusion}

Vibrato matching can be used to first suppress existing vibrato in a signal and then transfer vibrato from another signal.
Previously proposed methods for vibrato suppression are sufficient, but extensions to previous methods for vibrato transfer are necessary for vibrato matching.
While the prior vibrato transfer algorithm uses a simple AM envelope that is applied to the entire signal, in vibrato matching different AM patterns are applied per-harmonic and across the spectral envelope of the residual signal.
This method of AM transfer, while more complex, is consistent with the complex nature of AM in natural vibrato~\cite{Roebel2011,Zhang2015}.
The source code for vibrato matching and original audio files are available online on the supplemental website\footnote{
\url{https://jeremyhyrkas.com/ICMC2026}
}, as well as additional figures and audio examples from Sections~\ref{sec:modifying}--~\ref{sec:blending}.

Vibrato matching can be used as a method for modifying vibrato in a signal in isolation.
Use cases include changing existing vibrato by example, transferring vibrato between instrument classes to create instrument sounds not possible acoustically, or augmenting physical modeling synthesis.
Matching modulation patterns between sources also offers unique creative opportunities when sources are mixed together.
While listening tests have not yet been conducted to confirm that using vibrato matching on unison signals reduces a listener's ability to perceive multiple sources,
source separation algorithms inspired by similar listening tests show a reduced capacity for separation after modulation patterns are matched.
Mixes of vibrato matched signals can be used to create hybrid instrument sounds from recordings of different instrument types, or for creative control of the perception of shared movement in a mix.

\begin{acknowledgments}
The author would like to thank Professors Tamara Smyth, Miller Puckette, and Tom Erbe for their invaluable insights.
Additional thanks to the DMA students in Performance at the Music Department of the University of California San Diego who generously provided examples of their playing with and without vibrato.
\end{acknowledgments} 

\vfill\break 
\bibliography{bib}

\end{document}